%
%

\input harvmac.tex

\def\IR{\relax{\rm I\kern-.18em R}}
\def\IZ{\relax\ifmmode\mathchoice
{\hbox{\cmss Z\kern-.4em Z}}{\hbox{\cmss Z\kern-.4em Z}}
{\lower.9pt\hbox{\cmsss Z\kern-.4em Z}}
{\lower1.2pt\hbox{\cmsss Z\kern-.4em Z}}\else{\cmss Z\kern-.4em
Z}\fi}
\font\cmss=cmss10 \font\cmsss=cmss10 at 7pt

\def\mnk{{\cal M}_{N,k}}
\def\mzeronk{{\cal M}_{N,k}^0}
\def\tmnk{{\tilde{\cal M}}_{N,k}}
\def\tX{{\tilde X}}
\def\tx{{\tilde x}}
\def\tq{{\tilde q}}
\def\cn{{\cal N}}

\def\np#1#2#3{Nucl. Phys. {\bf B#1} (#2) #3}
\def\pl#1#2#3{Phys. Lett. {\bf #1B} (#2) #3}
\def\prl#1#2#3{Phys. Rev. Lett. {\bf #1} (#2) #3}
\def\physrev#1#2#3{Phys. Rev. {\bf D#1} (#2) #3}

\def\cmp#1#2#3{Comm. Math. Phys. {\bf #1} (#2) #3}

\def\ptp#1#2#3{Prog. Theor. Phys. {\bf #1} (#2) #3}

\lref\abkss{O. Aharony, M. Berkooz, S. Kachru, N. Seiberg, and E. 
Silverstein, 
``Matrix Description of Interacting Theories in Six Dimensions,'' 
hep-th/9707079.}%

\lref\abks{O. Aharony, M. Berkooz, S. Kachru, and E. 
Silverstein, 
``Matrix Description of $(1,0)$ Theories in Six Dimensions,'' 
hep-th/9709118.}%

\lref\lowe{D. A. Lowe, ``$E_8\times E_8$ Small Instantons in Matrix
Theory,'' hep-th/9709015.}%

\lref\casher{A. Casher, ``Gauge Fields on the Null Plane,''
\physrev{14}{1976}{452}.}%

\lref\thorn{C. B. Thorn, ``Quark Confinement in the Infinite Momentum
Frame,'' \physrev{19}{1979}{639}\semi
C. B. Thorn, ``A Fock Space Description of the $1/N_c$ Expansion of
Quantum Chromodynamics,'' \physrev{20}{1979}{1435}\semi
C. B. Thorn, ``Asymptotic Freedom in the Infinite Momentum Frame,''
\physrev{20}{1979}{1934}.}%

\lref\natidlcq{N. Seiberg, ``Why is the Matrix Model Correct ?,''
\prl{79}{1997}{3577}, hep-th/9710009.}

\lref\masyam{T. Maskawa and K. Yamawaki, ``The Problem of $P_+=0$ Mode
in the Null Plane Field Theory and Dirac's Method of Quantization,''
\ptp{56}{1976}{270}.}%

\lref\dlcq{H. C. Pauli and S. J. Brodsky, ``Solving Field Theory in
One Space One Time Dimension,'' \physrev{32}{1985}{1993}\semi
H. C. Pauli and S. J. Brodsky, ``Discretized Light Cone Quantization :
Solution to a Field Theory in One Space One Time Dimensions,''
\physrev{32}{1985}{2001}.}%

\lref\berdoug{M. Berkooz and M. Douglas, ``Five-branes in M(atrix) 
Theory,''
\pl{395}{1997}{196}, hep-th/9610236.}%

\lref\adhm{M. Atiyah, V. Drinfeld, N. Hitchin, and Y. Manin,
``Construction of Instantons,'' \pl{65}{1978}{185}.}

\lref\various{E. Witten, ``String Theory Dynamics in Various 
Dimensions,''
hep-th/9503124, \np{443}{1995}{85}.}

\lref\nahm{W. Nahm, ``Supersymmetries and Their Representations,''
\np{135}{1978}{149}.}

\lref\minwalla{S. Minwalla, ``Restrictions Imposed by Superconformal Invariance on Quantum Field Theories,'' hep-th/9712074.}

\lref\distler{J. Distler, private communication.}

\lref\nakajima{H. Nakajima, ``Resolutions of Moduli Spaces of Ideal 
Instantons on $\IR^4$,'' in ``Topology, Geometry and Field Theory,''
ed. Fukaya, Furuta, Kohno and Kotschick, World Scientific.}

\lref\yamron{J. P. Yamron, ``Topological Actions from Twisted
Supersymmetric Theories,'' \pl{213}{1988}{325}.}

\lref\vafwit{C. Vafa and E. Witten, ``A Strong Coupling Test of S
Duality,'' hep-th/9408074, \np{431}{1994}{3}.}

\lref\wittenstring{E. Witten, ``On the Conformal Theory of the Higgs
Branch,'' hep-th/9707093.}%

\lref\dougmoore{M. R. Douglas and G. Moore, ``D-branes, Quivers and
ALE Instantons,'' hep-th/9603167.}%

\lref\dfromm{P. K. Townsend, ``D-branes from M-branes,''
hep-th/9512062, \pl{373}{1996}{68}.}

\lref\fivebranes{
O. Aharony, ``String Theory Dualities from M Theory,'' hep-th/9604103,
\np{476}{1996}{470}\semi
E. Bergshoeff, M. de Roo and T. Ortin, ``The Eleven Dimensional
Five-brane,'' hep-th/9606118, \pl{386}{1996}{85}\semi
P. Pasti, D. Sorokin and M. Tonin, ``Covariant Action for a $D=11$
Fivebrane with the Chiral Field,'' hep-th/9701037,
\pl{398}{1997}{41}\semi
I. Bandos, K. Lechner, A. Nurmagambetov, P. Pasti, D. Sorokin and
M. Tonin, ``Covariant Action for the Superfive-brane of M Theory,''
hep-th/9701149, \prl{78}{1997}{4332}\semi
M. Aganagic, J. Park, C. Popescu and J. H. Schwarz, ``World Volume
Action of the M Theory Fivebrane,'' hep-th/9701166,
\np{496}{1997}{191}\semi
I. Bandos, N. Berkovits and D. Sorokin, ``Duality-Symmetric
Eleven-Dimensional Supergravity and its Coupling to M-branes,''
hep-th/9711055.}%

\lref\yoshioka{K. Yoshioka, ``The Betti Numbers of the Moduli Space of
Stable Sheaves of Rank 2 on ${\bf P}^2$,'' J. Reine Angew. Math. {\bf
453} (1994) 193, Proposition 5.4.}%

\lref\gny{I. Grojnowski, H. Nakajima and K. Yoshioka, in preparation.}%

\lref\malstr{J. Maldacena and A. Strominger, ``Semi-classical Decay 
of Near Extremal Five-branes,'' hep-th/9710014.}%

\lref\sanjohn{J. Brodie and S. Ramgoolam, ``On Matrix Models of M5
Branes,'' hep-th/9711001.}%

\lref\moorelitt{A. Losev, G. Moore, and 
S. Shatashvili, ``M$\&$m's,'' hep-th/9707250.}

\lref\natistring{N. Seiberg, ``New Theories in Six Dimensions and
Matrix Description of M-theory on $T^5$ and $T^5/\IZ_2$,''
hep-th/9705221.}%

\lref\helpol{S. Hellerman and J. Polchinski, ``Compactification in 
the Lightlike Limit,'' hep-th/9711037.}%

\lref\bfss{T. Banks, W. Fischler, S. Shenker, and L. Susskind, 
``M theory as a Matrix Model:  A Conjecture,'' hep-th/9610043,
\physrev{55}{1997}{112}.}%

\lref\susskind{L. Susskind, ``Another Conjecture About Matrix Theory,''
hep-th/9704080.}%

\lref\future{O. Aharony, M. Berkooz and N. Seiberg, work in progress.}

\lref\forms{E. Witten, ``Constraints on Supersymmetry Breaking,''
\np{202}{1982}{253}\semi
D. Friedan and P. Windey, ``Supersymmetric Derivation of the
Atiyah-Singer Index and the Chiral Anomaly,'' \np{235}{1984}{395}\semi
L. Alvarez-Gaume, ``Supersymmetry and the Atiyah-Singer Index
Theorem,'' \cmp{90}{1983}{161}}%

\lref\disagree{
M. R. Douglas, H. Ooguri and S. H. Shenker, ``Issues in M(atrix) Model
Compactification,'' \pl{402}{1997}{36}, hep-th/9702203\semi
M. Dine and A. Rajaraman, ``Multigraviton Scattering in
the Matrix Model,'' hep-th/9710174\semi
M. R. Douglas and H. Ooguri, ``Why Matrix Theory is Hard,''
hep-th/9710178\semi
D. Kabat and W. Taylor, ``Spherical Membranes in Matrix Theory,''
hep-th/9711078.}%

\lref\banksreview{T. Banks, ``Matrix Theory,'' hep-th/9710231.}%

\lref\bigsus{D. Bigatti and L. Susskind, ``A Note on Discrete Light
Cone Quantization,'' hep-th/9711063.}%

\lref\review{S. J. Brodsky, H. Pauli and S. S. Pinsky, ``Quantum
Chromodynamics and Other Field Theories on the Light Cone,''
hep-ph/9705477.}%

\lref\wittentwoz{E. Witten, ``Some Comments on String Dynamics,''
hep-th/9507121, Contributed to STRINGS 95: 
Future Perspectives in String Theory, Los Angeles, CA, 13-18 Mar 
1995.}

\lref\brs{
M. Rozali, ``Matrix Theory and U Duality in Seven Dimensions,''
hep-th/9702136, \pl{400}{1997}{260}\semi
M. Berkooz, M. Rozali and N. Seiberg,  ``Matrix Description
of M theory on $T^4$ and $T^5$,'' hep-th/9704089, \pl{408}{1997}{105}.}%

\lref\strominger{A. Strominger, ``Open P-branes,'' hep-th/9512059,
\pl{383}{1996}{44}.}

\lref\wittennone{E. Witten, ``Branes and the Dynamics of QCD,''
hep-th/9706109.}%

\lref\susskindint{L. Susskind, unpublished.}%

\lref\ndouglas{M. R. Douglas, ``Branes Within Branes,'' hep-th/9512077.}

\lref\silwit{E. Silverstein and E. Witten, ``Criteria for Conformal
Invariance of $(0,2)$ Models,'' hep-th/9503212, \np{444}{1995}{161}.}%

\lref\juan{J. Maldacena, ``The Large $N$ Limit of Superconformal Field
Theories and Supergravity,'' hep-th/9711200.}%

\lref\kallosh{P. Claus, R. Kallosh and A. Van Proeyen, ``M 5-brane and
Superconformal $(0,2)$ Tensor Multiplet in 6 Dimensions,''
hep-th/9711161.}%

\lref\cds{A. Connes, M. R. Douglas and A. Schwarz, ``Noncommutative
Geometry and Matrix Theory: Compactification on Tori,'' hep-th/9711162.}%

\lref\orisav{O. J. Ganor and S. Sethi, ``New Perspectives on Yang-Mills 
Theories with Sixteen Supersymmetries,'' hep-th/9712071.}%

\lref\savbound{S. Sethi and M. Stern, 
``D-Brane Bound States Redux,'' hep-th/9705046\semi
P. Yi, ``Witten Index and Threshold Bound States of D-Branes,''
\np{505}{1997}{307}, hep-th/9704098\semi 
M. Porrati and A. Rozenberg,
``Bound States at Threshold in Supersymmetric Quantum Mechanics,''
hep-th/9708119.}

\lref\opr{N. A. Obers, B. Pioline and E. Rabinovici, ``M Theory and U
Duality on $T^d$ with Gauge Backgrounds,'' hep-th/9712084.}%

\lref\dwpp{B. de Wit, K. Peeters and J. C. Plefka, ``Supermembranes
and Supermatrix Models,'' hep-th/9712082.}%



\Title{\vbox{\baselineskip12pt\hbox{hep-th/9712117}
\hbox{IASSNS-HEP-97/126, RU-97-93}}}
{\vbox{\centerline{Light-Cone Description of $(2,0)$ Superconformal}
\centerline{}
\centerline{Theories in Six Dimensions}}}

\centerline{Ofer Aharony$^1$, Micha Berkooz$^2$, and Nathan Seiberg$^2$} 
\smallskip
\smallskip
\centerline{$^1$ Department of Physics and Astronomy}
\centerline{Rutgers University }
\centerline{Piscataway, NJ 08855-0849, USA}
\centerline{\tt oferah@physics.rutgers.edu}
\smallskip
\smallskip
\smallskip
\centerline{$^2$ Institute for Advanced Study}
\centerline{Princeton, NJ 08540, USA}
\centerline{\tt berkooz@ias.edu, seiberg@ias.edu}
\bigskip
\bigskip
\noindent
We study the $(2,0)$ superconformal theories in six dimensions, which
arise from the low-energy limit of $k$ coincident 5-branes, using
their discrete light-cone formulation as a superconformal quantum
mechanical sigma model. We analyze the realization of the
superconformal symmetry in the quantum mechanics, and the realization
of primary operators. As an example we compute the spectrum of chiral
primary states in symmetric $Spin(5)_R$ representations. 
To facilitate the analysis we introduce and briefly discuss
a new class of Lorentz non-invariant theories, which flow in the IR to
the $(2,0)$ superconformal field theories but differ from them in the UV.

\Date{December 1997}


\newsec{Introduction}

Last year, a proposal for a light-cone quantization (or DLCQ) of M
theory in terms of a quantum-mechanical system, called Matrix theory,
appeared in \bfss. The DLCQ (Discrete Light-Cone Quantization
\refs{\masyam,\casher,\thorn,\dlcq})
interpretation of this proposal appeared in \susskind, and a
derivation of the conjecture was given in
\natidlcq. The idea that the large $N$ Matrix theory provides a
light-cone Hamiltonian for M theory has passed many tests. However,
the status of the conjecture that the finite $N$ theory provides a
DLCQ of M theory is less clear, since direct comparisons of the
low-energy limit of the finite $N$ theory with supergravity do not
always work \disagree. There is no compelling argument why a direct
comparison with supergravity should work for finite $N$, since the
low-energy limit of a DLCQ of M theory does not have to be equivalent
to a DLCQ of supergravity \refs{\helpol,\banksreview}. 

A simpler arena to analyze this question may be in DLCQ descriptions
of field theories (see \review\ for a review of works on this subject,
and \refs{\helpol,\bigsus} for recent works motivated by Matrix
theory).  The simplest case of such a description is the DLCQ proposed
for $(2,0)$ superconformal theories in six dimensions (which were
discovered in
\refs{\wittentwoz,\strominger}). The DLCQ description involves a
quantum-mechanical sigma model, and it was derived from Matrix
theory in \refs{\abkss,\wittenstring}. A generalization of this
description to $(1,0)$ theories in six dimensions appeared in
\refs{\lowe,\abks}, but we will not discuss it here.

In this paper we begin a detailed exploration of the DLCQ description
of the $A_{k-1}$ $(2,0)$ superconformal theories in six dimensions,
which arise from the low-energy limit of $k$ coincident 5-branes. One
motivation for this study is the need to better understand the
properties of Matrix-like DLCQ descriptions. The other motivation is
the study of the $(2,0)$ superconformal theories in six dimensions,
which are the simplest examples of non-trivial field theories above
four dimensions. They are also interesting for applications to Matrix
theory (see, e.g., \brs) and to field theory (it was conjectured in
\wittennone\ that these theories may be related to the large $N$ limit
of QCD). According to a conjecture of \juan, they may also be related
to M theory on $AdS_7\times S^4$.

In section 2 we describe the light-cone (and DLCQ) construction of
$(2,0)$ field theories, and give a new derivation of it along the lines of
\natidlcq. In section 3 we analyze how the superconformal algebra
looks in the light-cone frame, and see how it is realized in the
quantum-mechanical description.  The quantum mechanical description
involves a sigma model on a singular space, and it seems that we need
to resolve the singularities in order to be able to make sense of the
model. In section 4 we describe such a resolution and its
interpretation in terms of the space-time theory.  In section 5 we
analyze general properties of superconformal theories in DLCQ.  In
section 6 we compute the spectrum of chiral primary operators which
are in symmetric representations of the $Spin(5)$ R-symmetry, and
interpret the chiral primary fields in terms of the natural
coordinates parametrizing the moduli space of these theories. Section
7 includes a detailed analysis of the behavior of such chiral primary
operators in some simple examples.  We briefly outline a procedure to
calculate $n$-point functions. The analysis of other primary
operators, more complicated OPEs and explicit computations of higher
$n$-point functions is left to future work \future.

As this paper was being completed, some overlapping results pertaining
to the (free) theory of a single 5-brane appeared in \sanjohn, and a
discussion of the superalgebra and of the single 5-brane theory
appeared also in \kallosh. For a related discussion of the
$(2,0)$ theories compactified on tori, see \orisav.
  
\newsec{The Quantum-Mechanical Light-Cone Description}

In \abkss, a quantum-mechanical model was conjectured to give a
light-cone (or DLCQ) description of the six dimensional field theory
corresponding to the low-energy theory of $k$ M
theory 5-branes. This includes the $A_{k-1}$ $(2,0)$ superconformal
theory as well as a decoupled free tensor multiplet. In this section
we will review this model and its derivations, before going on to
using it as a description of the spacetime theory in the rest of the
paper.

\subsec{A Direct Derivation of the DLCQ Description}

In DLCQ, a light-like coordinate (which we will choose to be $x^- =
{1\over 2}(x^0 - x^1)$) is compactified on a circle of radius $R$, and
the time coordinate is taken to be $x^+ = {1\over 2}(x^0 + x^1)$. The
compactification of a light-like circle may be viewed as a limit of
compactifications of near light-like circles, which in turn are
equivalent (by a Lorentz boost) to compactifications of standard space
coordinates. The light-like limit is obtained by taking the space-like
circle to be very small, and looking at the theory of the modes which
carry momentum around this circle. In general, one must also be
careful to keep modes which correspond to finite energies in the
original theory \natidlcq. In our case this is relatively simple since
the original theory we are starting with is conformal.

Thus, in order to obtain a DLCQ of the $(2,0)$ SCFTs we should look at
their compactification on a circle of radius $R_s$ and take the limit
$R_s \to 0$. At energies below the scale $1/R_s$, the compactified
theory is a $U(k)$ five dimensional $\cn=2$ Super Yang-Mills theory,
with a coupling constant $g_5^2 = R_s$, which goes to zero in the limit
we are interested in. Note that this is very different from the
behavior of standard field theories, which become strongly coupled
when compactified on small circles, causing problems in their direct
analysis in this way \helpol.

After the compactification, the momentum modes around the circle
become instanton-like particles of the SYM theory \various\ (namely,
particles which are charged under the global current $J = *(F \wedge
F)$). Finite energies in the original theory translate into very small
velocities for these particles. Thus, in the limit $R_s \to 0$, the
theory of $N$ of these modes, corresponding to a DLCQ with momentum
$P_-=N/R$, reduces to a quantum mechanics on the moduli space of
$N$ $U(k)$ instantons embedded in $\IR^4$, as in \abkss. We will
denote this moduli space by $\mnk$. For large $N$, this gives a
light-cone description of the uncompactified $(2,0)$ theory.

This provides an alternative derivation of the light-cone (or DLCQ)
description of \abkss. However, the resulting theory we find here is
not obviously well-defined, since the moduli space of instantons has
singularities corresponding to small instantons. These singularities
are not described just by the five dimensional $\cn=2$ SYM theory,
since this description breaks down at short distances (the theory is
non-renormalizable). Thus, we need to add some information on how to
regularize the singularities in the instanton moduli space. Another
apparent problem is that the moduli space we find is non-compact (in
directions that do not correspond to space-time, in addition to the
obvious non-compactness of the directions which are identified with
space-time). This was interpreted in \abkss\ as related to the IR
behavior of the conformal theory, and this will be exemplified in
detail in the following.

\subsec{A Derivation from M Theory with 5-branes}

One possible regularization is provided by the construction of the
quantum-mechanical model as a limit of a DLCQ description of M theory
\refs{\bfss,\susskind} with 5-branes, as described in \abkss. One
starts with a complete description of M theory in the presence of $k$
5-branes \berdoug, which is given by a quantum mechanical $U(N)$ SQCD
theory with 8 (real) supercharges, with one adjoint hypermultiplet and
$k$ hypermultiplets in the fundamental representation \ndouglas. For
finite gauge coupling $g_{QM}$, this is a non-singular theory which is
well-defined. The limit in which gravity decouples from the 5-brane
theory ($M_p \to \infty$) corresponds to the $g_{QM} \to
\infty$ limit of the quantum mechanical gauge theory, in which the
Coulomb branch decouples from the Higgs branch. All the massive modes
decouple in this limit, and the theory of the 5-branes is described by
a supersymmetric sigma model on the Higgs branch of this
theory\foot{Note that since this is a conformal theory, we do not need
to assume that the low-energy limit of the DLCQ of M theory is the
same as the DLCQ of supergravity.}. It is
a well-known fact \adhm\ that this Higgs branch is exactly the moduli
space of instantons $\mnk$ described above. In fact, the gauge theory
gives the simplest construction of this moduli space, called the ADHM
construction. It also provides for us a regularization, in the sense
that we can, in principle, compute any space-time correlation function
in the quantum mechanics for finite $g_{QM}$, and then take the limit
$g_{QM} \to \infty$ and get the correlation functions of the
superconformal theory. However, this regularization adds many more
degrees of freedom than we actually need (in particular, it adds all
of eleven dimensional supergravity). We will describe in section 4 a
different regularization which is more useful for our purposes.

\subsec{A Description of the Model}

Let us now give a more concrete description of our theory. As we
mentioned, the simplest description of the moduli space $\mnk$ is as
the Higgs branch of a $U(N)$ gauge theory with 8 supercharges, an
adjoint hypermultiplet and $k$ fundamental hypermultiplets. Let us
denote the scalar components of these hypermultiplets by $X$,$\tilde
X$ (two complex scalars in the adjoint representation of $U(N)$) and
by $q_i$,${\tilde q}^i$ in the $\bf N$ and $\bf{\overline{N}}$
representations of $U(N)$ ($i=1,\cdots,k$). The Higgs branch is
parametrized by the values of these fields, subject to the constraints
enforcing the vanishing of the scalar potential :
\eqn\Dconstraints{[X,X^\dagger] - [\tX,\tX^\dagger] + q_i q_i^\dagger
- (\tq^i)^\dagger (\tq^i) = 0}
and
\eqn\Fconstraints{[X,\tX] + q_i \tq^i = 0,}
and modded out by the $U(N)$ gauge symmetry. The total (real) dimension
of this space is $4Nk + 4N^2 - 3N^2 - N^2 = 4Nk$, and it is a
hyperK\"ahler manifold. 

The equations (and gauge symmetry) do not act on 4 decoupled
coordinates which are $\tr(X)$,$\tr(\tX)$. Thus, the moduli space
decomposes as $\mnk = \IR^4 \times \mzeronk$. The decoupled $\IR^4$
part will give rise to eight non-linearly realized supersymmetries,
which act by shifting the corresponding fermions. It will sometimes be
convenient to denote the coordinates of the $\IR^4$ component by
$\tilde{x}$, and those of $\mzeronk$ by $v$.

There is a natural metric on the moduli space, which is the
hyperK\"ahler metric. From the gauge theory point of view this is the
classical metric on the Higgs branch, which is not renormalized. On
the total space of fields, we can define a scalar function
\eqn\kahler{K = |q_i|^2 + |\tq^i|^2 + |X|^2 + |\tilde X|^2,}
given by the sum of the absolute value squared of all scalar
fields. The restriction of this function to the moduli space defines a
scalar function on the moduli space, which we will also denote by
$K$. Then, the metric on the moduli space is $g_{ij} = \half \del_i
\del_j K$ where $i,j$ label the coordinates of the moduli space. Note
that the space is scale-invariant.

The global symmetries of the theory are $SU(2)_R \times SU(2)_L \times
Spin(5) \times U(k)$. The first two factors correspond to the rotation
symmetries inside the 5-brane transverse to the light-cone coordinate,
the third factor is the rotations transverse to the 5-brane (or,
equivalently, the
R-symmetry of the $\cn=(2,0)$ supersymmetry in spacetime), and the last
factor corresponds to the gauge symmetry of the $(2,0)$ theory after
it is compactified on a circle. The supercharges in the quantum
mechanics are in the $\bf(2,1,4,1)$ representation of this group, so
that $SU(2)_R\times Spin(5)$ is the R-symmetry of the quantum mechanics.
The fields described above are
in the following representations :
\eqn\representations{\matrix{
&U(N)&SU(2)_R&SU(2)_L&Spin(5)&U(k)\cr
X_H&\bf{N^2}&\bf{2}&\bf{2}&\bf{1}&\bf{1}\cr
\Theta_X&\bf{N^2}&\bf{1}&\bf{2}&\bf{4}&\bf{1}\cr
q_H&\bf{N}&\bf{2}&\bf{1}&\bf{1}&\bf{k}\cr
\psi_q&\bf{N}&\bf{1}&\bf{1}&\bf{4}&\bf{k},\cr}}
where $X_H$ denotes the scalars in the adjoint hypermultiplets ($X$
and $\tX$), $q_H$ denote the scalars in the fundamental
hypermultiplets ($q$ and $\tq$), 
$\Theta_X$ are the fermionic partners of $X,\tilde{X}$, $\psi_q$
are the fermionic partners of $q,\tilde{q}$, and all the fields obey
appropriate reality conditions.
The bosonic coordinates are all neutral under the $Spin(5)$,
while the fermionic coordinates are all in the $\bf{4}$
representation. The other global symmetries 
act on the moduli space in a non-trivial way (except for $SU(2)_R$
they commute with supersymmetry).

\newsec{The Superconformal Algebra in Spacetime and in the Quantum 
Mechanics}

\subsec{Bosonic Components in the Spacetime Algebra}

The bosonic part of the superconformal algebra in six dimensions includes
the $SO(6,2)$ conformal algebra
\eqn\sosixtwo{[M_{\alpha\beta},M_{\gamma\delta}]=
-i(\eta_{\alpha\gamma}M_{\beta\delta}+\eta_{\beta\delta}M_{\alpha\gamma}
-\eta_{\alpha\delta}M_{\beta\gamma}-\eta_{\beta\gamma}M_{\alpha\delta}),}
where $\alpha,\beta=0,\cdots,7$, $\eta_{\alpha\beta}={\rm 
diag}(-1,1^6,-1)$.

The components $M_{ij}$ of this algebra ($i,j=0,\cdots,5$) are
identified with the usual Lorentz generators, while the other
components are related to the standard conformal generators by
\eqn\identpkd{P_i=M_{6i}+M_{7i}; \qquad K_i=M_{6i}-M_{7i}; \qquad
D=M_{67}.}
Here $P_i$ is the translation generator, $K_i$ generates special
conformal transformations and $D$ generates dilatations. With these
identifications, \sosixtwo\ leads to the usual conformal
algebra 
\eqn\suconf{\eqalign{[M_{ij},P_k]=-i(\eta_{ik}P_j-\eta_{jk}P_i); &\qquad
[M_{ij},K_k]=-i(\eta_{ik}K_j-\eta_{jk}K_i); \cr
[M_{ij},M_{kl}]=-i\eta_{ik}M_{jl} \pm permutations; &\qquad
[M_{ij},D]=0; \qquad
[D,K_i]=iK_i; \cr
[D,P_i]=-iP_i; &\qquad
[P_i,K_j]=-2iM_{ij}+2i\eta_{ij}D. \cr}}

Our description of this theory in discrete light cone quantization
sees only a sector of the theory with fixed $P_- = P_0 - P_1 = N/R$.
Thus, in the quantum mechanical description we should only be able to
see those elements of the superconformal algebra that commute with
$P_-$. These include the Galilean generators $M_{ij}$, $P_i$,
$H=P_+=P_0+P_1$ and $V_i = M_{0i}-M_{1i}$ (where now
$i,j=2,3,4,5$). Two other elements of the conformal algebra also
commute with $P_-$: they are $K_- = K_0 - K_1$ and $T = D - M_{01}$. 
$T$ has a natural interpretation as a dilatation followed by a
boost in the $x^1$ direction, which is needed to cancel the effect of
the dilatation on $P_-$.

The non-zero commutation relations of these elements are :
\eqn\subconf{\eqalign{[M_{ij},P_k]=-i(\eta_{ik}P_j-\eta_{jk}P_i); &\qquad
[M_{ij},V_k]=-i(\eta_{ik}V_j-\eta_{jk}V_i); \cr
[P_i,V_j]=-i\eta_{ij}P_-; \qquad [T,P_i]=-iP_i; &\qquad 
[T,V_i]=iV_i; \qquad
[P_i,K_-]=2iV_i; \cr [H,V_i]=-2iP_i; \qquad
[H,K_-]=-4iT; &\qquad [T,H]=-2iH; \qquad [T,K_-]=2iK_-. \cr}}

Note the explicit appearance of $P_-$ in the commutation relations as
a central term. The analysis of the bosonic part of the superconformal
algebra does not depend on the dimension of space-time, and we expect
to find the same algebra \subconf\ in a DLCQ
description of any conformal theory. Of course, the fermionic part
described below will depend strongly on the fact that we are in $5+1$
dimensions.

\subsec{Fermionic Components of the Superconformal Algebra}

The $(2,0)$ superconformal algebra in six dimensions 
includes 32 fermionic generators, which are all
$SO(6,2)$ spinors of the same chirality (say, $\bf 8_s$) and are a
fundamental $\bf{4}$ of $Sp(2)\simeq Spin(5)$. Their commutation relation
is of the form
\eqn\qcomm{\{Q_\alpha^i,Q_\beta^j\}= J^{ij}
(\Gamma^{\mu\rho})_{\alpha\beta}
M_{\mu\rho}+\delta_{\alpha\beta}R^{ij},}
where $J$ is the anti-symmetric form of $Sp(2)$, $R$ is the charge of
the $Sp(2)$ current, and
$\Gamma^{\mu\rho} = \half [\Gamma^{\mu},\Gamma^{\rho}]$.

It will be convenient to decompose the $Q$ operators according to
their charges under the $SO(1,1)$ symmetries rotating $x^0,x^1$ and
$x^6,x^7$, or, equivalently, according to the eigenvalues of $\Gamma^0
\Gamma^1$ and $\Gamma^6 \Gamma^7$ in the spinor representation of
$SO(6,2)$. Defining
\eqn\gammas{\Gamma^\pm={1\over 2}(\Gamma^0\pm\Gamma^1);
\qquad {\hat\Gamma}^\pm={1\over2}(\Gamma^6\pm\Gamma^7),}
the decomposition into eigenstates $Q,\tilde{Q},S$ and $\tilde{S}$ 
satisfies
\eqn\decomposition{\matrix{
          &\Gamma^0\Gamma^1&\Gamma^6\Gamma^7\cr
S         &1              &1              &
\Gamma^+S={\hat\Gamma}^+S=0\cr
{\tilde S}&-1               &1              &\Gamma^-{\tilde S}=
					  {\hat\Gamma}^+{\tilde S}=0\cr
Q         &-1               &-1             &
\Gamma^-Q={\hat\Gamma}^-Q=0\cr
{\tilde Q}&1              &-1             &\Gamma^+{\tilde Q}=
	{\hat\Gamma}^-{\tilde Q}=0.\cr
}}

Then, the commutation relations \qcomm\ are schematically of the form
(neglecting the R charge contributions, and without writing down the
indices, which follow from the symmetries): 
\eqn\fermcomm{\eqalign{
&\{Q,Q\} \sim H; \qquad \{Q,\tilde{Q}\} \sim P_i; \qquad
\{\tilde{Q},\tilde{Q}\} \sim P_-; \cr
&\{Q,\tilde{S}\} \sim M_{0i}+M_{1i}; \qquad 
\{\tilde{Q},S\} \sim V_i; \cr
&\{Q,S\} \sim M_{ij}^R - T; \qquad \{\tilde{Q}, \tilde{S}\} 
\sim M_{ij}^L - (M_{01}+D); \cr
&\{S,S\} \sim K_-; \qquad \{S,\tilde{S}\} \sim K_i; \qquad
\{\tilde{S},\tilde{S}\} \sim K_+, \cr}}
where we have decomposed the $SO(4)$ generators $M_{ij}$ into
$SU(2)_R$ generators $M_{ij}^R$ and $SU(2)_L$ generators $M_{ij}^L$.

\subsec{Identification of the Bosonic Elements in the Quantum
Mechanics}

Let us now try to identify the algebra \subconf\ in our quantum
mechanical description. This description includes four variables
$\tilde{x}^i$ which are free and decoupled, corresponding to the
center of mass position. The other $(4Nk-4)$ variables $v^k$ are
coordinates on a non-trivial hyperK\"ahler manifold. The metric on the
target space of the quantum mechanical sigma model is of the form
$g_{i j} =
\half \del_i \del_j K$ where the function $K$ was described in the
previous section (the inverse metric will be denoted by $g^{kl}$,
$g^{kl} g_{lj} = \delta^k_j$). We will denote all $4Nk$ coordinates by 
the
collective name $x^i = \{ \tilde{x}^j, v^k\}$. The conjugate momenta
to these variables will be denoted by $\Pi_{x^i}$ ($[\Pi_{x^i},x^j] =
-i\delta_i^j$). Their action on wave functions is the same as a
derivative, $\Pi_x \sim -i\del_x$. 

The identification of the Galilean subalgebra of the superconformal
algebra is mostly straightforward. The operator $H$ is just the
Hamiltonian of the quantum mechanical sigma model
\eqn\Hamiltonia{H = {R\over N} (\delta_{ij} \Pi_{\tilde{x}^i}
\Pi_{\tilde{x}^j} +
g^{kl} \Pi_{v^k} \Pi_{v^l}) + fermions.}
Note that we chose the Hamiltonian to be explicitly proportional to
$R/N$. This is the natural scaling for the interpretation of this
Hamiltonian as the DLCQ of a space-time theory, since then $H = P_+ =
(P_i^2 + M^2) / P_- = R(P_i^2 + M^2) / N$. The natural scaling for the
quantum-mechanical sigma-model Hamiltonian does not include this
factor of $R/N$. However, since the sigma model is scale invariant, the
two conventions differ only by rescaling all coordinates and momenta by
$\sqrt{R/N}$. In particular, correlation functions in the sigma model
will be functions of $\sqrt{N\over R}\cdot x$, 
and not of $x$ and $R/N$ separately.

The momenta $P_i$ and boosts $V_i$ act only on the center of mass
coordinates, so we can identify them with
\eqn\momenta{P_i = \Pi_{\tilde{x}^i}; \qquad V_i = P_- \tilde{x}^i.}
The rotations $M_{ij}$ generate an $SO(4) \simeq SU(2)_R \times
SU(2)_L$ global symmetry in the quantum mechanics, which acts in a
rather complicated way on the sigma model coordinates. 
These operators satisfy the commutation relations \subconf.

Next, we should identify the operators $T$ and $K_-$. For this we
note that (using \subconf) we may identify $H$, ${1\over 2}T$ and
${1\over 4}K_-$ with generators of an $SO(1,2)$ algebra.
We can identify this algebra with the
conformal algebra of the quantum mechanical theory\foot{Using
the same conventions as \sosixtwo\ and \identpkd\ for the conformal
algebra of the quantum mechanics, $H \sim (P_0)_{QM}$, $T \sim 2D_{QM}$
and
$K_- \sim 4(K_0)_{QM}$.},
generated by 
$\delta t = \epsilon_1 + \epsilon_2 t + \epsilon_3 t^2$.
In the case at hand, we may explicitly write
these as
\eqn\qmconf{\eqalign{T &= -(\half g^{ij} {\del K \over \del x^i}
\Pi_{x^j} - 2iNk) =
-{1\over 4} g^{ij} ({\del K \over \del x^i}
\Pi_{x^j} + \Pi_{x^i} {\del K \over \del x^j}) \cr
K_- &= - {1\over 4} 
P_- g^{ij} {\del K \over \del x^i} {\del K \over \del x^j}, \cr}}
up to additional fermionic terms. The expressions above are meaningful
only away from the singularities of the manifold, where additional
contributions may be localized.
In the conformal quantum mechanics,
these operators obey the commutation relations \subconf. Note that
$K_-$ vanishes only at the origin of the moduli space ($x=0$).

Obviously, these identifications are meant to be relevant only at some
specific time, say $t=0$. The operators may then be evolved in time in
the usual way, by the equation ${\cal O}(t) = e^{itH} {\cal O}(0)
e^{-itH}$. For instance, $V_i(t) \simeq P_- \tilde{x}^i + 2t
\Pi_{\tilde{x}^i}$.  Note that we expect to find $H = P_+ = ((P_i)^2 +
M^2) / P_-$, so we may identify the second ($v$-dependent) part of $H$
with $M^2$ in spacetime (which is a Casimir operator of the Lorentz
algebra but not of the conformal algebra).

\subsec{Identification of the Fermionic Operators in the Quantum 
Mechanics}

It is easy to identify also most of the operators described in
\S3.2 in
the quantum mechanical description. The charges of the $Sp(2)$
R-currents in spacetime may be identified with the $Sp(2)$ R-charges
of the quantum mechanics, which involve only the fermions.
Using the commutation relations \fermcomm\ it is easy to check that $Q$,
$\tilde{Q}$ and $S$ commute with $P_-$, while $\tilde{S}$ does not, so
we do not expect to identify $\tilde{S}$ 
in the quantum mechanical description.

The operators $Q$ and $\tilde{Q}$ are simply the spacetime
supersymmetry generators. The $Q$'s are the generators which are
linearly realized in the quantum mechanics as the 8 supercharges of
the quantum mechanics. Schematically they are of the form $Q^\alpha_a
\sim \sqrt{R/N} \Theta_a^i \Pi_{x_\alpha^i}$, where we divided the
target space coordinates into $SU(2)_R$ doublets.  $\tilde Q$ is
non-linearly realized in the quantum mechanics, and schematically it
is given by $\tilde{Q}^{\dot{\alpha}}_a \sim 2\sqrt{P_-}
\Theta^{\dot{\alpha}}_a$, where $\Theta^{\dot{\alpha}}_a$ is the
superpartner of $\tilde{x}^i$ (denoted by $\tr(\Theta_X)$ above).
$\tilde Q$ is in the $\bf(1,2,4,1)$ representation of the global
symmetry, while $Q$ and $S$ are in the $\bf(2,1,4,1)$ representation.

The superconformal generator $S$ may be identified in the quantum
mechanics (up to a constant) with the other fermionic generator of the
quantum mechanical superconformal algebra. Schematically, it is given
by $S^\alpha_a \sim \sqrt{P_-}
\Theta_a^i {\del K \over \del x_\alpha^i}$. It is easy to check that with
these identifications, the commutation relations \fermcomm\ are obeyed.
The $Q$'s and $S$'s form doublets of the $SO(1,2)$ algebra
mentioned above ($[T,Q] \sim -iQ, [T,S] \sim iS, [K_-,Q]
\sim -iS, [H, S] \sim -iQ, [K_-,S] = [H,Q] = 0$). 

To summarize, 
we have found that our quantum mechanical sigma model is actually a
superconformal sigma model. This is consistent with the fact that the
R-symmetry group is $SU(2)_R\times Spin(5)$, which is one of the
possible R-symmetries for superconformal quantum mechanics \nahm. The
superconformal algebra of the quantum mechanical theory includes the
generators $H=P_+$, $T$, $K_-$, $Q$ and $S$ of the space-time
superconformal algebra, all of which commute with $P_-$. The other
generators which commute with $P_-$ are realized either as charges of
the $SU(2)_R\times SU(2)_L\times SO(5)$ global symmetries, or (in the
case of $P_i$, $V_i$ and $\tilde{Q}$) as simple operators acting on
the decoupled free fields corresponding to the center of mass
position.

\newsec{Resolution of the singularities}

\subsec{The need for the resolution}

It is natural to organize the states of the space-time superconformal
theory into representations of the superconformal algebra. The
representations which appear in physical theories include primary
states, which are annihilated by all the special conformal generators
$K_\mu$ and $S$ (we will elaborate more on this in section 5). Since
we identified $K_-$ with the special conformal generator of the
quantum mechanical theory, such states will correspond to primary
states in the quantum mechanics\foot{States that are not primary
in space-time may also appear as primary in the quantum mechanics,
since the $0+1$ dimensional 
superconformal algebra is smaller.}, which are annihilated by
$K_-$. 

 From the form of $K_-$ given above, it is clear that such a state
must be concentrated completely at the origin of moduli space. Other
local operators, obtained by acting with $P$ and $Q$ on the primary
operator, will also be concentrated there. The fact that local
operators in space-time correspond to states in the quantum mechanics
which are localized at the origin of the moduli space had been
anticipated by L. Susskind
\susskindint\ based on the following intuitive picture. We are 
interested in local operators in space-time, which are a disturbance
only in a very small region of space-time. Thus, their wave function
should certainly be localized in the quantum mechanical variables
which correspond to space-time positions. These arise from the $X$
fields in the gauge theory description. The other variables in the
quantum mechanics, arising from the $q$ fields, are associated with
the instanton size. Large instantons are not expected to
correspond to local operators, so we expect states corresponding to
local operators to be localized at $q=\tq=0$.

Using the algebra, one can arrive at the same conclusion in another
way. The way to obtain a local operator in space-time is to create a
disturbance in a compact region of space-time and then shrink it using
the dilatation operator. The corresponding action in the quantum
mechanics will be to shrink all the support of any wave function to
the origin.

There is one obvious wave function of this type, which is just
$\prod_i \delta(x^i)$, but since the origin is very singular it is
hard to say if there are not more states that are ``hidden inside the
singularities.'' Thus, we would like to be able to resolve the 
singularities and deal with a smooth manifold.
Luckily,
there exists a simple resolution of the space which makes it
completely smooth.
In this section
we will describe this resolution and its physical interpretation in
space-time.

As described in section 2, the space $\mnk$ is the Higgs branch of a
gauge theory with 8 supercharges, which is the space of $X$'s and
$q$'s subject to \Dconstraints\ and \Fconstraints\ and modded out by
the gauge group. This space is singular whenever the gauge group is
not completely broken, since then there are additional massless fields
in the gauge theory. From the point of view of the instanton moduli
space these singularities correspond to small instantons. By adding a
Fayet-Iliopoulos term to the gauge theory, we can force the gauge
group to be completely broken at all points on the Higgs branch, and
then the Higgs branch is no longer singular. The Fayet-Iliopoulos
parameter consists of 3 real scalars, which are a triplet of
$SU(2)_R$. They appear on the right side of the equations
\Dconstraints\ and \Fconstraints. By an $SU(2)_R$ transformation we
can always
choose just one of these scalars to be non-zero. The Higgs branch of
the theory, which is the deformed moduli space $\tmnk$, is then given
by the space of $X$'s and $q$'s subject to the equations
\eqn\nDconstraints{[X,X^\dagger] - [\tX,\tX^\dagger] + q_i q_i^\dagger
- (\tq^i)^\dagger (\tq^i) = \zeta I_N}
and
\eqn\nFconstraints{[X,\tX] + q_i \tq^i = 0,}
modded out by the gauge group,
where $\zeta$ is some non-zero real number and $I_N$ is the $N\times
N$ unit matrix.

 From the mathematical viewpoint turning on $\zeta$ is a
compactification of the small-instanton region of the moduli
space. This compactification was used in \vafwit\ to compute partition
functions of $\cn=4$ 4D gauge theories, and in \nakajima\ to compute
the cohomology of $\mnk$. It has no known interpretation in terms of
instantons, but it is natural in the context of their generalization
to stable sheaves in algebraic geometry.

\subsec{Spacetime Interpretation of the Resolution}

What is the physical interpretation of the parameter $\zeta$ in the
space-time theory ? We can discuss this question within the theory of the
5-branes alone, or we can do so within the $U(N)$ quantum
mechanics corresponding to the full M theory with the 5-branes, 
which flows in the IR (on one of its branches) to the
5-brane theory. We will use the latter, more general, description.
Apriori, not every deformation of the quantum mechanics corresponds
to a change in the theory in space-time, as the space-time
interpretation may collapse. We do not expect this, however, to happen
in our case. 

Strictly speaking, the deformation by $\zeta$ lifts the Coulomb
branch, which corresponds to 0-branes moving away from the
longitudinal 5-brane. It does so, however, in a very limited
way. Unlike its pronounced effect on the origin of the Higgs branch
and of the Coulomb branch, its effect far out along the Coulomb
branch is merely an addition of a constant to the DLCQ
Hamiltonian. As such, it changes the dynamics there
only by an insignificant phase. We therefore still expect
to have, for example, bound states corresponding to gravitons
\savbound. We have not lost the space-time interpretation, and the
deformation by $\zeta$ should have such an interpretation.

We will argue that the space-time interpretation of the parameter
$\zeta$ is that of turning on a constant 3-form field $C$ in M
theory. This field is not gauge invariant, and can be gauged away in
the absence of 5-branes. However, in the presence of a 5-brane, gauge
transformations of the $C$ field in the bulk act also on the self-dual
3-form field strength $H$ living on the 5-brane, and the gauge
invariant (self-dual) field on the 5-brane is actually $(H-C)$
\dfromm. Thus, by a gauge transformation we can turn the 3-form field
$C$ in the bulk into a constant (self-dual) 3-form field strength $H$
on the 5-brane.  The component of the $H$ field that we turn on is
$H_{+ij}$, where the $i,j$ indices are in the non-compact directions
of the 5-brane and the $+$ index is the DLCQ time direction. In most
of the analysis below we will use this gauge freedom and discuss the
space-time theory with an $H$-field turned on (we can only do this
explicitly for the case of a single 5-brane, but this should not
change our results).

To get the precise coupling we can follow the procedure of
\natidlcq. We will do so with some detail.  At the first step we are
interested in going from M theory on an almost light-like circle,
which we will denote by $M$, to M theory on a space-like circle with
radius $R_s$, a configuration which we will denote by $M_s$ (we will
label quantities in $M_s$ by a subscript $s$). We will then go to
the $\widetilde M$ system by an appropriate rescaling.

In the original $M$ system, on a light-like circle, we wish to set
$H_{+ij}$ to some non-zero value. The DLCQ Hamiltonian is of the
general form
\eqn\hamm{P_+=RM_p^2F\biggl({P_i\over M_p},{H_{+ij}\over RM_p^4}\biggr),} 
for some function $F$.
This form is determined on dimensional grounds, and by the requirement
that the DLCQ Hamiltonian be covariant under longitudinal
boosts. Under a longitudinal boost the dynamical quantities change
according to their $M_{01}$ quantum numbers, but we also change the
value of $R$. The quantity $H_{+ij}/ R$ remains invariant under
such a boost, and, therefore, it is the correct one to enter the
function $F$. The factors of $M_p$ are then fixed by
dimensionality. Our final goal is to go over to the $\widetilde M$
system in a way that preserves the physics, as encoded by $F$.

Next, as in \natidlcq, we make the circle slightly space-like and then
boost the system to the configuration $M_s$, in which M theory is
compactified on a spatial circle with radius $R_s$. When we deform the
circle to be slightly space-like we do not change $H$ and preserve the
fact that $H_{-ij}=0$. The reason we must do this is that the
self-duality condition on the 3-form field $H$ implies that the
physical components of $H_{+ij}$ transform as a triplet of $SU(2)_R$
(and a singlet of $SU(2)_L$), while those of $H_{-ij}$ transform as a
triplet under $SU(2)_L$ (and a singlet of $SU(2)_R$)\foot{Recall that
we decomposed the $SO(4)$ rotation symmetry of the 5-brane transverse
to the light-cone direction into $SU(2)_R\times SU(2)_L$.}. This is
true irrespectively of details of the circle on which we
compactify. As our deformation has (by construction) only $SU(2)_R$
quantum numbers, we need to maintain that $H_{-ij}=0$.

Now, when we go to the $M_s$ system after the boost, we find that
\eqn\resch{H_{s,+ij}={R_s\over R}H_{+ij},\ H_{s,-ij}=0,}
and that the Hamiltonian is now of the form
\eqn\hamms{(P_0)_s=R_sM_p^2F\biggl(
{P_i\over M_p},{H_{s,+ij}\over R_sM_p^4}\biggr).}

In order to go to $\widetilde M$, we now want to rescale time and
distances such that the energy of the physical processes we are
interested in remains fixed, and the relevant physics remains the
same. This criterion determines that, in going to $\widetilde M$, we
need to keep the arguments of the function $F$ fixed. In \natidlcq\
this determined the scaling ${P_i\over M_p}={{\tilde P}_i\over {\tilde
M}_p}$. Here it also determines the relation
\eqn\impresc{{H_{s,+ij}\over R_s M_p^4}=
{{\tilde H}_{+ij}\over R_s {\tilde M}_p^4}.}

Once we are in the $\widetilde M$ system, and have determined all the
quantities in this system, we can go over to the IIA description
(which is valid since the radius $R_s$ of the space-like circle in
this system is small) and write down the Lagrangian for the 0-branes.
${\tilde H}_{+ij}$ may now be interpreted as a non-zero field strength
on the 4-brane, with a specific chirality\foot{When we wrap the
5-brane on a circle of radius $R_s$ in the $x^1$ direction to get a
D4-brane \dfromm, the $4+1$ dimensional field strength on the D4-brane
is $F_{ij} = R_s H_{1ij}$.}. Such a field enters the Lagrangian of the
D0-branes as a FI term. This can be seen in several ways. For
instance, a computation in the free conformal field theory of the 0-4
strings shows that a field strength shifts the mass of the 0-4 strings
in the same way that a FI term does.

A more intuitive way is to perform a gauge transformation on the bulk
fields in the string theory, such that the constant field strength $F$
on the D4-brane becomes a constant NS B-field in the bulk (this is
similar to the inverse of the gauge transformation we used above to
turn $C$ into $H$).  The low-energy action of D-branes does not
contain just the $trF^2$ term, but rather $tr(F-B)^2$, where $B$ is
the pullback of the space-time NS 2-form field to the D-brane
worldvolume.  If we start with the action for a wrapped 2-brane and
T-dualize to go to the action of 0-branes, this becomes a term of the
form $tr([X_\mu,X_\nu]-B_{\mu\nu})^2$, and again we see that some of
the components of the $B$ field appear as a FI term.

Going back to the quantum mechanical Lagrangian for the D0-branes in
the $\widetilde M$ system, we find that it is of the schematic form
(suppressing indices)
\eqn\tildlag{{\tilde L}={1\over R_s}{\dot{\tilde X}}^2+
R_s{\tilde M_p}^6\biggl([{\tilde
X},{\tilde X}]+{{\tilde H}_+\over R_s{\tilde M_p}^6}\biggr)^2.}

Using the relation to the original M-theory variables \natidlcq\ 
\eqn\rescx{X=\biggl({R\over R_s}\biggr)^{1\over 2}{\tilde X},\
M_p^2R={\tilde M_p}^2R_s}
and
\eqn\finresc{{{\tilde H}_+\over
R_s{\tilde M_p}^4}={H_{s,+}\over R_sM_p^4}={H_+\over RM_p^4},} one
obtains
\eqn\finl{L={1\over R}{\dot X}^2+RM_p^6\biggl([X_i,X_j]+ {H_{+ij}\over
RM_p^6}\biggr)^2.}

Thus, we identify the Fayet-Iliopoulos term $\zeta$ with
$H_{+ij}\over R M_p^6$, or more generally with $C_{+ij} \over R
M_p^6$, in the space-time M theory that we are describing\foot{This
was recently noted also in \refs{\cds,\opr,\dwpp}.}.

Note that our identification of the Fayet-Iliopoulos term with the
$C_{+ij}$ field in space-time is completely general, and can be used
even in the absence of any 5-branes. Of course, in such a case this
field may locally be completely gauged away. This is
consistent with the fact that in the absence of any charged matter
($q$'s), the Fayet-Iliopoulos term only contributes a constant to the
Hamiltonian, which does not change the dynamics. In the presence of
5-branes and the constant $C$-field, we find that particles with
longitudinal momentum are attracted to the 5-branes, since the minimal
energy away from the 5-branes (where $q=0$) is $|\zeta|^2$, while
inside the 5-branes it is zero (for configurations satisfying
\nDconstraints\ and \nFconstraints).

\subsec{New Lorentz Non-Invariant Theories}

In the previous subsection we described the role of the FI term in the
full theory of M theory with $k$ 5-branes, described by the $U(N)$
gauge quantum mechanics. However, in this paper we are actually
interested in taking $M_p\rightarrow\infty$, in order to decouple the
$(2,0)$ field theory on the 5-brane from the bulk. Since, as mentioned
above and described in more detail below, we would like to use $\zeta$
as a UV cutoff for the quantum mechanical sigma model that we obtain,
we would like to take this limit such that the FI term remains
fixed. However, since we found above that $\zeta\simeq {C_{+ij}\over
RM_p^6}$, we need to take $C_{+ij}\propto M_p^6\rightarrow\infty$ in
this limit. Thus, the deformed quantum mechanical theory (the sigma
model on the deformed moduli space $\tmnk$) describes in space-time
the theory of $k$ 5-branes as $M_p \to \infty$, with a $C$-field which
also goes to infinity as $C \propto M_p^6$. Obviously, such a limit
is not accessible from the low-energy supergravity theory (for a
single 5-brane we can translate the $C$ field into an $H$ field whose
energy density is very large in Planck units), but in principle it
exists in M theory, and our DLCQ description provides a construction
of the resulting theory.

The resulting theory is not equivalent to the $(2,0)$ field theory, as
one clearly sees from the fact that its DLCQ description is quite
different (obviously, it is not Lorentz invariant, since we explicitly
broke the $SU(2)_R$ which was part of the space-time Lorentz
group). Many of the UV properties of the resulting theory will be
different from those of the $(2,0)$ theory (as one can compute
explicitly using the quantum mechanics), but it will still flow to the
$(2,0)$ theory in the IR.  This may be deduced from the fact that the
change induced by the FI term is particularly important near the
origin of the Higgs branch, and is less important far along the flat
directions (thus, the long distance behavior of the new theory is
governed by the $(2,0)$ fixed point). This also shows that it is
incorrect to think of this theory as a mere change of the vacuum of
the $(2,0)$ field theory, as this would not modify the theory in the
UV. Instead, one can think of this as a change in the short-distance
(UV) properties of this theory which does not change its long-distance
properties.

To summarize, our construction naturally implies the existence of a
new class of Lorentz non-invariant theories, whose UV properties are
very different from those of the standard field theories (the theory
is not Lorentz invariant nor scale invariant in the UV)\foot{Note that
these theories are not Lorentz invariant but still have 16
supercharges. They are similar in this sense to the theories recently
discussed in \cds.}. Even though we have not encountered such theories
before, and we do not seem to be able to access any of their
properties through the low-energy supergravity theory\foot{For $k=1$
and for values of $C$ (or $H$) that are small in Planck units, it
should be possible to explicitly compute the effect of turning on this
field also directly in space-time
\refs{\dfromm,\fivebranes,\kallosh}. However, it is not known how to
do this for $k>1$, and in any case we are interested in large values
of $C$ (or $H$) in Planck units, when the low-energy description is no
longer valid.}, the preceding construction is (up to possible
subtleties of the large $N$ limit) a complete calculable definition of
these theories.

\subsec{$\zeta$ as an Ultra-Violet Cutoff}

As an ultra-violet cutoff $\zeta$ appears in the theory in a familiar
and straightforward way. $\zeta$ has negative mass dimension $(-2)$,
so taking $\zeta$ to be small is equivalent to going to large time
differences.  When analyzing the theory with finite $\zeta$, we will
usually be interested only in the long-time behavior of correlation
functions.  At long times the quantum mechanical theory will flow to a
superconformal theory, which will contain the information about the
superconformal theory in space-time.

The procedure of extracting $n$-point functions from the theory with the
cut-off is standard. The key point is that we are interested in the
behavior only as $t\rightarrow\infty$. By appropriate wave-function
renormalization we will extract the quantities that remain finite as
we take the UV cut-off to infinity or, equivalently, flow to the
IR\foot{A similar normalization of wave functions to cancel the
dependence on a cutoff which is a FI term appears in \silwit.}. All
other quantities, with a non-trivial dependence on the cut-off, will
be interpreted as artifacts of the cut-off.

Let us briefly demonstrate this standard procedure for 2 and 3-point
functions. The $x$ dependence of wave functions can always be chosen to
be through the combination $x/\sqrt{\zeta}$. Then, the time dependence
will always appear through the combination $\zeta/Rt$.
Given a wave function $\phi_1({x/\sqrt{\zeta}})$, we
can expand its 2-point function as
\eqn\explarget{ \langle\phi_1(t)\vert\phi_1(0)\rangle 
\sim 
a \biggl( {\zeta \over Rt}\biggr)^{2d} + {\rm higher\ order\ 
terms\ in\ } 
{\zeta \over Rt},}
where $d$ is the dimension of $\phi_1$ in the quantum mechanics
(which, as described above, is proportional to $T$).
We then define a renormalized operator by $\phi_1\rightarrow
(\zeta/R)^{-d}a^{-{1/2}}\phi_1$. 
This wave function renormalization is
performed such that the coefficient of the identity operator in the
2-point function is exactly ${1\over t^{2d}}$. The terms of higher
order in ${1\over t}$ will have powers of $\zeta$ in them, and are,
therefore, cut-off artifacts.

For any choice of generic wave functions $\phi_1$ and $\phi_2$ with
the same quantum numbers (under global symmetries), the large-$t$
behavior of the 2-point function is dominated by the conformal state
of lowest dimension with the same quantum numbers, i.e, the two point
functions $\langle\phi_i(t)\vert\phi_i(0)\rangle,\ i=1,2$ decay (for
large $t$) with the same power of $t$. From these two functions,
however, we can extract a state that corresponds to an operator with
higher dimension by taking a linear combination $\phi_3$ such that the
large-$t$ behavior of $\langle \phi_3(t) \vert\phi_3(0) \rangle$ now
falls off with a higher power of ${1\over t}$. One again needs to
rescale this operator such that the leading contribution does not
depend on $\zeta$, and interpret all remaining $\zeta$ dependence as a
cut-off artifact. $\phi_3$ can now serve as a representative for the
next-to-leading operator with the same quantum numbers, as far as the
large-$t$ behavior is concerned.

Next, let us go over to the 3-point functions. Taking some 3 operators
$\psi_{1,2,3}$ and calculating their 3-point function,
one obtains an expansion
(before the wave function renormalization)
for $\langle\psi_1(t_1)\psi_2(t_2)\psi_3(0)\rangle$ of the form
\eqn\threepoint{c
\biggl({\zeta\over Rt_1}\biggr)^{d_1+d_3-d_2}
\biggl({\zeta\over Rt_2}\biggr)^{d_2+d_3-d_1} 
\biggl({\zeta\over R(t_1-t_2)}\biggr)^{d_1+d_2-d_3} + \cdots,}
where $(\cdots)$ denotes higher order terms in $\zeta/Rt$.
After the wave function
renormalization, the $\zeta$ dependence disappears from the leading
term, which becomes the 3-point function of the
operators. Subleading terms are, as before, cut-off artifacts.

To conclude, the definition of the states in the conformal field
theory, i.e. when $\zeta=0$, is as the Hilbert space of states on the
resolved target space $\tmnk$, modulo a relation. The relation is that
two such elements in the Hilbert space are equivalent if the leading
${1\over t}$ dependence of all their correlation functions are the
same. After this identification, we can translate the finite-$\zeta$
wave functions to the conformal states by the procedure described
above.

An example of this is the following. Let us consider the real axis,
and the evolution of functions by the heat equation. Upon Fourier
transform, the long time behavior of the wave function is governed by
the Taylor expansion of the Fourier transform around the origin $P=0$,
${\hat f}(p)=a_0+a_1p+{1\over2}a_2p^2+\cdots$. The equivalence class of
the state with the least dimension contains all the states with
$a_0\neq 0$, the equivalence class of the state with the
next-to-leading dimension is given by functions with $a_0=0,a_1\not=0$,
etc.

We cannot do such a computation explicitly in our case, as the space
is quite complicated. However, the general structure will be the
same. The Fourier transform will be replaced by the spectral
representation of the Schr\"odinger equation. The relevant components
of the spectrum will be either normalizable discrete zeros or a
continuum component that goes down to zero. Two functions will give
the same state in the conformal limit if their behavior at
eigenvalues close to zero is the same.

We will not analyze the general equivalence relation any further, as
we will be interested in a special class of operators and states that
are easier to control, which correspond to chiral primary fields. In
order to analyze these, we need to understand better the
supersymmetries of the resolved model.

\subsec{Supersymmetries and Forms in Quantum Mechanical Sigma Models}

After the resolution of the singularities, we have a quantum
mechanical sigma model with 8 supercharges on a smooth (though
non-compact) manifold. The theory after the resolution is no longer
superconformal. In this subsection we describe a convenient
description of the fermionic variables and supersymmetries of such
models in terms of forms on the target space.

Let us begin by assuming that we have a quantum mechanical
supersymmetric sigma model with two supercharges (or one complex
supercharge $Q$) on a manifold $\cal M$. Then, it is well known
\refs{\forms} that the states of the sigma model may be identified
with forms on $\cal M$, and one of the supercharges may be
identified with the exterior
derivative operator $d$ acting on these forms. The
SUSY algebra acts as $[Q,x^i] = \theta^i$ (where $\theta^i$ is a
complex fermion), and the fermionic zero modes obey
$\{(\theta^i)^\dagger,\theta^j\} = g^{ij}$ where $g$ is the metric on
the space. We can choose the vacuum such that $\theta_j|0\rangle = 0$,
and then we can identify the fermion creation operators
$(\theta^i)^\dagger$ with forms $dx^i$, and the annihilation operators
$\theta^i$ with $*dx^i*$. This allows us to translate any state in the
quantum mechanics into a form on the target space. The SUSY generator
$Q \sim \theta^\dagger_i \del_i +
\Gamma^i_{jk}(x) \theta_i
\theta^\dagger_j \theta^\dagger_k$ may then be identified with the
exterior derivative operator $d$ on the space (the other SUSY
generator corresponds to $d^\dagger=*d*$). Thus, the space of states
which are annihilated by $Q$ is translated into the space of closed
forms on the target space $\cal M$.

The $\cn=2$ SUSY algebra includes a $U(1)_R$ symmetry, which $Q$ and all
the fermionic fields are charged under. The identification above means
that the $U(1)_R$ charge of a state is equal to the degree of the
corresponding form, up to a constant shift
(the $U(1)_R$ charge of a $p$-form state is $p-{1\over 2}{\rm dim}(\cal
M)$).

In our case we actually have 8 (real) supercharges, and the R symmetry
is $SU(2)_R\times Spin(5)$. The $SU(2)_R$ is broken by $\zeta$ to
$U(1)_\zeta$. Choosing an $\cn=2$ subalgebra includes a choice of a
$U(1)_R$ symmetry in $U(1)_\zeta \times Spin(5)$. Since we want a 
$U(1)_R$
symmetry that does not act on the bosonic coordinates, we should
choose $U(1)_R
\subset Spin(5)$, and it seems that the unique consistent choice of a
$U(1)_R$ is to use the $U(1)$ which appears in the maximal subgroup of
$Spin(5)\simeq Sp(2)$ which is $SU(2)\times U(1)$. In particular, the 
$\bf
4$ of $Spin(5)$ decomposes as $\bf{2}_1 + \bf{2}_{-1}$, so with this
choice of $U(1)_R$ we are assured that all the fermionic coordinates
will have charge $\pm 1$, as required for the realization of the SUSY
algebra described above. This choice is also required to ensure that the
fermions transform as the tangent bundle of the manifold, as needed
for their identification with forms.

After this choice of an $\cn=2$ subalgebra of the quantum mechanical
supersymmetry, one of the $Q$ operators with charge $(+1)$ under the
$U(1)_R$ symmetry will be the $\cn=2$ supersymmetry generator, which
may be identified with the exterior derivative operator $d$ (its
conjugate is identified with $d^\dagger$).  The other components of
$Q$ correspond to other differential operators on $\cal M$, of degrees
$\pm 1$.

\newsec{Representations of the Conformal Algebra and Correlators in DLCQ}

\subsec{Representations of the Superconformal Algebra in DLCQ}

The interesting representations of the conformal group for physical
applications are representations containing a primary field $\Phi$.
The action of the conformal generators on such a field is given by
\eqn\primary{\eqalign{
[P_\mu, \Phi(x)] &= i\del_\mu \Phi(x) \cr
[M_{\mu \nu}, \Phi(x)] &= [i(x_\mu \del_\nu - x_\nu \del_\mu) +
\Sigma_{\mu \nu}] \Phi(x) \cr
[D, \Phi(x)] &= i(d+x^\nu \del_\nu) \Phi(x) \cr
[K_\mu, \Phi(x)] &= [-i(x^2 \del_\mu - 2x_\mu x^\nu \del_\nu - 2x_\mu
d) + 2 x^\nu \Sigma_{\mu \nu}] \Phi(x), \cr}}
and, in particular, $[K_\mu, \Phi(0)] = 0$. $d$ is called the
dimension of the field $\Phi$, and $\Sigma_{\mu \nu}$ are the usual
spin representations of the Lorentz group.

In the usual construction of representations of the conformal group,
$D$ is diagonalized. The primary field (at $x=0$) is annihilated by
the generators $K_\mu$ of the conformal group which lower the
dimension $d$, while the action of the other conformal generators on
the primary field gives us the full conformal block. In particular,
the fields fall into representations of the Lorentz group and of any
other global symmetry, and the generators of these groups commute with
$D$. For the superconformal group the story is similar, with
the $S$ generators of the superconformal group annihilating
primary states, while the supersymmetry generators $Q$ raise the value
of $d$ and generate the conformal block.

In the discrete light-cone quantization, the representation structure
is different since we are looking only at a sector with a particular
eigenvalue of $P_-$. In particular, this sector may include a
particular momentum mode of all the fields of the conformal block
described above. Since, as discussed above, $D$ does not commute with
$P_-$, we cannot diagonalize $D$, and instead we diagonalize $T = D -
M_{01}$. We will call the eigenvalue of $T$ (divided by $i$)
the ``DLCQ dimension'' of a field, and denote it by $t_0$. We also do
not see full representations of the Lorentz group, but only of the
part of the group that commutes with $P_-$, as described above. In
particular, for the $(2,0)$ superconformal theories in six dimensions,
we will sort the states according to the eigenvalue of $T$ and the
$SU(2)_R\times SU(2)_L\times Spin(5)$ representation.

As before, we will look for states which are annihilated by the
generators of the conformal group which lower the ``DLCQ dimension''
of a field. These generators are $K_-$, $V_i$ and $S$. Given such a
state, we can form its full conformal block by acting on it with the
other superconformal generators. The generators $\tilde{Q}$ and the
$SU(2)_R\times SU(2)_L\times Spin(5)$ charges commute with $T$, and
generate states with the same ``DLCQ dimension.'' The other
operators $Q$, $P_i$ and $H$, raise the ``DLCQ dimension'' of
a field, and generate various descendants of a primary field.

As an example let us analyze the free tensor multiplet. It contains
five scalar fields, in the $\bf(1,5)$ representation of $SO(5,1)\times
Spin(5)$, one fermion field in the $\bf(4,4)$ representation and one
self-dual 3-form field strength in the $\bf(10,1)$
representation. Since these are all free fields, their dimensions are
$d=2,5/2$ and $3$ respectively. The full representation of the
superconformal algebra is completed by additional operations with $Q$
and $P$ on these fields.

In DLCQ, we can easily compute the $T$-eigenvalues and $SU(2)_R\times
SU(2)_L\times Spin(5)$ representations of all of these fields. The 5
scalars all have a ``DLCQ dimension'' $t_0=2$, and are in the
$\bf(1,1,5)$ representation. The fermions split according to their
$M_{01} \sim \Gamma^0 \Gamma^1$ eigenvalue, into a $\bf(1,2,4)$
representation with $t_0=2$ and a $\bf(2,1,4)$ representation with
$t_0=3$. The Dirac equation for a free fermion of momentum $P_-=N/R$
allows us to express the $t_0=3$ components as derivatives
(descendants) of the $t_0=2$ components.
The tensor field splits into a $\bf(1,3,1)$ with $t_0=2$, a
$\bf(2,2,1)$ with $t_0=3$ and a $\bf(3,1,1)$ with $t_0=4$. 
The $t_0=3$ and $t_0=4$ states are again descendants, in the DLCQ,
of the $t_0=2$ states.
The $t_0=2$ states of the tensor multiplet
form an irreducible representation 
of the Clifford algebra of the $\tilde Q$ operators.
Generally, to generate all the states 
we will need to also use the $SU(2)_R\times
SU(2)_L\times Spin(5)$ charges. 

A special class of primary fields is the class of chiral primary
fields. These are representations of the superconformal group that
contain null states. We will discuss here only chiral primary fields
for which some combination of supercharges $Q$ acting on the primary
states in the representation (states with the minimal dimension)
vanishes. We will further restrict ourselves to representations whose
primary fields are Lorentz scalars. The superconformal algebra may be
used to derive a bound on the dimension of primary fields, given their
R-symmetry representation, which is exactly saturated for chiral
primary fields. The free tensor multiplet described above is the
simplest case of a chiral primary field.

We will not derive here the general equation for the dimension of
chiral primary fields \refs{\minwalla,\distler}, but only a special
case that will be useful in the following. The $(2,0)$ superconformal
algebra in six dimensions contains as a subalgebra the $(1,0)$
superconformal algebra. The R-symmetry of the $(1,0)$ algebra is
$SU(2)$, and this $SU(2)$ may be chosen to be either of the two
$SU(2)$ factors in the maximal subgroup $SU(2)\times SU(2) \subset
Spin(5)_R$ of the R-symmetry group of the $(2,0)$ algebra. The $(1,0)$
superconformal algebra may be used to show that the bound on the
dimension of a primary scalar field in the $2j+1$ dimensional
representation of the $SU(2)$ R-symmetry is $d \geq 4j$, and it is
saturated only for chiral primary fields of the $(1,0)$ superconformal
group. The $\bf{5}$ representation of $Spin(5)$ decomposes into
$\bf(2,2)+\bf(1,1)$ of $SU(2)\times SU(2)$, so we find that the bound
on a scalar field in the $\bf{5}$ representation is at least $d \geq
2$ (since obviously a primary field of the $(2,0)$ algebra is also a
primary field of the $(1,0)$ theory). In this case this bound is
actually the maximal one, as the example of the free tensor multiplet
demonstrates. Similarly, one can easily show that the dimension of a
primary scalar field in the $n$-th symmetric product of $\bf{5}$'s
obeys $d \geq 2n$, since its decomposition into $SU(2)\times SU(2)$
representations includes $\bf(n+1,n+1)$, and a field which saturates
this bound is necessarily a chiral primary field (since it is a chiral
primary field of the $(1,0)$ subalgebra).

\subsec{2-point Functions of Primary Operators in DLCQ}

The correlation functions of scalar primary fields are strongly
constrained by the conformal algebra. Their 2-point functions are of
the form
\eqn\twopoint{\langle\Phi_i^\dagger(x) \Phi_i(0)\rangle = 
{1\over (x^2)^{d_i}},}
and their 3-point functions are of the form
\eqn\threepoint{\langle\Phi_i(x_1) \Phi_j(x_2) \Phi_k(x_3)\rangle = 
{c_{ijk} \over
{
[(x_{12})^2]^{(d_i+d_j-d_k)/2}
[(x_{13})^2]^{(d_i+d_k-d_j)/2}
[(x_{23})^2]^{(d_k+d_j-d_i)/2}}},}
where $d_i$ is the dimension of $\Phi_i$ and $x_{ij} = x_i - x_j$.

In the DLCQ, we are not studying the operators $\Phi$, but
rather their momentum modes
\eqn\state{\Phi_N(x^i,x^+) = \int_0^{2\pi R} 
dx^- e^{-iNx^-/R} \Phi(x^i,x^+,x^-).}  We are interested in computing
things like the 2-point function of these momentum modes,
$f_N(x^i,x^+) = \langle \Phi^\dagger_N(x^i,x^+) \Phi_N(0) \rangle$.
The remaining conformal symmetry (which commutes with $P_-$) is still
enough to determine this up to a constant. The rotation symmetries
require $f_N(x^i,x^+) = f_N((x^i)^2,x^+)$. The action of $T$ then
requires $f_N(x^i,x^+) = {1\over (x^+)^d} {\hat
f}_N((x^i)^2/x^+)$. Since the operator \state\ is an eigenstate of
$T$, and $T$ is the same algebra element for every $N$, $d$ does not
depend on $N$.  Then, the action of $K_-$ uniquely determines
\eqn\corrdlcq{\langle \Phi^\dagger_N(x^i,x^+) \Phi_N(0) \rangle \propto
{1\over (x^+)^{d}} e^{-iN(x^i)^2 / 4Rx^+}.}

In particular, we can determine the dimension of primary operators in
the DLCQ without having to go to large values of $N$. For higher
$n$-point functions we may have to go to large $N$ to get exact
results in the space-time theory.

\newsec{Primary States and Correlation Functions from the Quantum 
Mechanics}

The remaining obstacle to calculating dimensions and correlation
functions is the precise identification between states in space-time
and states in the quantum mechanics.  As explained in section 4,
states of the conformal field theory in space-time (without the
cutoff) are realized in the quantum mechanics as equivalence classes
in the Hilbert space of the sigma model on the resolved space
$\tmnk$. We defined these equivalence classes using the long time
evolution of wave functions, but actually computing the equivalences
is quite complicated, and we will do it for some simple examples
below.

\subsec{Identifying Chiral Primary Operators in the Quantum Mechanics}

As described above, any primary operator in the quantum
mechanics must be annihilated by $K_-$. Chiral primary operators (of
the space-time theory) are also annihilated by some of the spacetime
SUSY generators (in addition to being annihilated by $K_-$). 
As described above, the states that
are annihilated by one particular $Q$, which generates 
(together with $Q^\dagger$) an $\cn=2$
subalgebra, correspond to closed forms on the target space. We will
focus here on this particular type of chiral primary states.

For our purposes, we are interested in states that are annihilated by
$Q$ and are also concentrated at the origin of the moduli space. Since
there is an obvious scaling symmetry of the target space (for
$\zeta=0$), we can scale any state which has compact support (namely,
it vanishes sufficiently fast on the boundary at infinity of the
manifold) so that it is concentrated at the origin of moduli
space. Thus, the chiral primary operators of an $\cn=2$ superconformal
sigma model are given by the cohomology with compact support
(sometimes denoted by $H(\mnk,\del
\mnk)$) of the target space (which, in our case, is the instanton
moduli space $\mnk$). 
Note that we can also use the $\cn=2$ quantum mechanical
superconformal algebra to learn about the
dimensions of chiral primary operators.
The $\cn=2$ supersymmetry algebra contains a $U(1)_R$ symmetry,
and the superconformal algebra gives a bound on the dimensions of
primary operators, $T \geq |R|$, with equality only for chiral primary
operators (we normalize the R-charge of $Q$ to be one, and $T\sim
2D_{QM}$ as described above).

The justification for identifying states differing by an exact form is
the following. Adding an exact form to a state certainly changes the
state, and also, for instance, its 2-point functions. This can be
identified with adding the commutator of $Q$ with some operator to our
primary state. However, such an exact form will always have a higher
``DLCQ dimension,'' because it will not saturate the bound relating
the dimension to the R-charge. Thus, its correlation functions will
decay faster in time, and the long-time behavior will not depend on
such additions of exact forms. The large-time correlation functions
depend only on the cohomology, which is the reason why we can identify
the cohomology with (a subgroup of) the chiral primary states.

In our case we actually have 8 supersymmetry generators, and we
are looking for states that are annihilated by some combination of the
supersymmetry generators. 
The simplest way to look for chiral primary operators is to take an
$\cn=2$ subalgebra of the $\cn=8$ SUSY algebra. Given such a choice of
an $\cn=2$ SUSY generator $Q$, the states annihilated by $Q$ will
correspond, as described above, to the cohomology with compact support
of $\mnk$. Obviously, a primary state that is annihilated by the
$\cn=2$ SUSY generator corresponds to a chiral primary operator, since
this generator is a combination of the $\cn=8$ SUSY generators.

However, the converse is not necessarily true.  There could be
representations which are chiral primaries of the $\cn=8$ algebra but
which do not include any chiral primaries of the $\cn=2$ subalgebra. A
chiral primary of the $\cn=8$ superconformal algebra, in some
representation of $SU(2)_R\times Spin(5)$ and with some ``DLCQ
dimension'' $t_0$, includes states of different $U(1)_R$
charge\foot{$t_0$ is determined in terms of the $SU(2)_R\times
Spin(5)$ representation by the $\cn=8$ superconformal algebra.}. If
the representation includes a state with $R = t_0$, this state will be
a chiral representation also of the $\cn=2$ subalgebra, and we will
find it in the cohomology. On the other hand, if the highest $U(1)_R$
charge appearing in the decomposition is lower than this bound, the
decomposition of this $\cn=8$ representation into $\cn=2$
representations will not include any chiral primary operators, so we
will not find a representative of this state in the cohomology. For
scalars in space-time, which are hence $SU(2)_R$ singlets, it is easy
to check that the only $Spin(5)$ representations which obey this
condition are the totally symmetric (traceless in each pair of
indices) products of $\bf{5}$'s. Thus, only chiral primary operators
in totally symmetric (traceless) representations will correspond to
the cohomology with compact support of $\mnk$. Other chiral primary
operators will be annihilated by other linear combinations of $Q$'s,
and we leave their analysis to future work
\future.

Note that the derivation above proves that the bound on the dimension
of primary operators in the quantum mechanics 
which are singlets of $SU(2)_R$ and
which are in the $n$-symmetric traceless representation of $Spin(5)$
is $T \geq 2n$. Using the relation to the space-time theory, it also
shows that the bound on the dimension of scalar primary operators in
$(2,0)$ six dimensional SCFTs in the same representation of
$Spin(5)_R$ is $d \geq 2n$, with equality exactly for chiral
primary operators, as we found above by using a $(1,0)$ subalgebra.
This is a special case of the general correspondence between
R-symmetry representations and dimensions of primary operators.

\subsec{Computation of the Cohomology}

Now that we have resolved the singularities, the computation of the
cohomology with compact support of the resolved space $\tmnk$ becomes
a well-defined mathematical problem. An algorithm for computing the
homology using Morse theory was given in \nakajima, and an explicit
formula appears in \yoshioka\ (a different proof will appear in
\gny). After translating the formula from homology to cohomology with
compact support by the usual duality, it states that a generating
function for the dimension $a_{p,N}$ of $H^p(\tmnk)$\foot{$H^p(\tmnk)$,
the space of $p$-forms in the cohomology, is isomorphic to
$H_{4Nk-p}(\tmnk)$, which is the space of ($4Nk-p$)-cycles in the
homology.} is
\eqn\formula{\prod_{i=1}^\infty \prod_{j=1}^k {1\over {1 - t^{2j}
q^i}} = \sum_{N,p} a_{p+2Nk,N} t^p q^N.}

Let us begin by interpreting this result for $N=1$. In this case, we
find a one dimensional cohomology for $p=2k+2,2k+4,\cdots,4k$. Using
the relation between the $R$-charge in the $\cn=2$ superconformal
algebra and the degree of the form (which is $R = p-{1\over 2} {\rm
dim}(\mnk) = p-2Nk$), this
implies the existence of $\cn=2$ superconformal chiral primary states
with $R=2,4,\cdots,2k$. As described above, this implies the existence
of states with $t_0=2,4,\cdots,2k$, where the state with $t_0=2n$ is in a
$Spin(5)$ representation corresponding to the $n$'th symmetric
(traceless) product of $\bf 5$'s. Note that an important consistency
check on this procedure is that \formula\ never predicts states with
negative dimension (or even states whose dimension is smaller than 2).

The states we found for $N=1$ have a natural interpretation as the
coordinates on the moduli space of the space-time theory, which is
$\IR^{5k}/S_k$. At a generic point on the moduli space, the low-energy
field content is $k$ tensor multiplets, each including 5 scalars
$\phi^i_a,\ i=1,\cdots,5,\ a=1,\cdots,k$, which are the natural local
coordinates on the moduli space. It is natural to look at (globally
well-defined) functions on the moduli space which are the symmetric
products of these scalar fields.  The symmetric products of more than
$k$ fields are determined by the symmetric products with less
fields\foot{Note that even the symmetric products with $k$ fields or
less are not completely independent.}. We find that for $N=1$ we have
exactly one state corresponding to the momentum one component of each
such symmetric product (of $k$ fields or less), and no other state in
a symmetric $Spin(5)$ representation. We will denote the operator in
the $l$-symmetric product of $\bf 5$'s by $u_l$. A special case is
$u_1$ which has dimension 2, and is, therefore, a free field (this can
be proven by using the superconformal algebra). This is exactly the
free decoupled tensor multiplet, whose scalars correspond to the
center of mass position of the $k$ 5-branes.

Given this field content, we would expect to have states for arbitrary
$N$ of the form $\prod_i (u_{l_i})_{N_i}$ where $(u_l)_n$ is the
$n$-momentum mode of $u_l$, $1\leq l_i \leq k$, the $Spin(5)$
representations are multiplied symmetrically\foot{Note that there are
no singularities in symmetric products of chiral primary fields in
symmetric representations, since the dimension of the product is equal
to the sum of the dimensions of the fields. Generally, this would not
be true, so there is no chiral ring for $(2,0)$ superconformal
theories.}, and $\sum_i N_i = N$. It is easy to compute the
contribution of these states to the generating function \formula, with
each such state appearing at a value of $p$ corresponding to a ``DLCQ
dimension'' $t_0=\sum_i 2l_i$, and we find that it is exactly equal to
\formula. Thus, it seems that all chiral primary states in 
symmetric $Spin(5)$ representations are of this form. Specifically,
the only chiral primary operators in symmetric (traceless)
representations of $Spin(5)$ are products of the $u_l$ fields, and a
product of $j$ $u_l$ fields appears for the first time at momentum
$N=j$. Note that apriori we do not know at which momentum a particular
field will first appear. Since our computation of the spectrum of
chiral primary operators in symmetric representations is valid for all
values of $N$, it is valid also in the large $N$ limit, which
corresponds to a light-cone description of the uncompactified six
dimensional theory.

For $k=1$ this result was described already in \abks, from a slightly
different point of view. In this case the theory is free, and the
states we find here correspond exactly to the chiral primary states of
the conformal theory of a free tensor field. In particular, from the
analysis of the free field theory we find that all the chiral primary
states are in totally symmetric (traceless) representations (since
anti-symmetric products vanish), so our construction gives us all the
chiral primary states in this case.

\newsec{Some Simple Examples}

\subsec{States and Correlation Functions for $N=1$ $k=1$}

In this section we will give some concrete examples of the general
construction of chiral primary states described in the previous
section. The simplest example is $N=1$ $k=1$. The theories with $k=1$
(corresponding to $U(1)$-instantons, which are pointlike) 
are always expected to be free.
The case $N=1$ is particularly simple, since the instanton moduli
space in this case is just $\IR^4$. In this case the deformation by
$\zeta$ has no effect, and we can do all our computations without
it. Still, it is worthwhile to analyze this case in some detail, both
because it is the simplest case, and because the moduli space always
includes a decoupled $\IR^4$ factor, and the behavior of the states in
this $\IR^4$ component will generally be very similar to the behavior
for $N=1$ $k=1$.

Using the formula \formula, we find that in this case there is only
one form in the cohomology with compact support, with degree 4. It is
dual to the only element in the homology, which is the point. An
element of this cohomology may be chosen to be $\hat{f} = 
f(x^1,x^2,x^3,x^4)
dx^1 \wedge dx^2 \wedge dx^3 \wedge dx^4$ for any function $f$ which
has compact support and obeys $\int_{\IR^4} {\hat f} \neq 0$. Then, it
is easy to explicitly check that this state is in the cohomology with
compact support, namely that $d\hat{f} = 0$ and that there is no 
$\hat{g}$
with compact support such that $\hat{f}=d\hat{g}$. As described above,
to get the actual chiral primary state we have to scale this state to
the origin, so that it is annihilated by $K_-$ (which in this case is
proportional to $|x|^2$). The resulting state is thus 
\eqn\fgadol{F = \prod_i
\delta(x^i) dx^1 \wedge dx^2 \wedge dx^3 \wedge dx^4,} up to a
normalization constant which we will determine by computing the
2-point function below\foot{Note that here we choose a different
convention than in \S4.4, in which the wave function depends on $x$
and not on $x/\sqrt{\zeta}$.  Of course, after the wave function
normalization we will get the same results.}. According to the general
arguments above, this state represents one of the scalar fields of the
free tensor multiplet $u_1$ with one unit of longitudinal momentum.

It is very simple in this case to find the other states with $t_0=2$
in the free tensor multiplet, since they may be generated by acting
with $\tilde{Q}$ on the state we found. In this case the $\tilde{Q}$
are just the fermionic partners of the $\IR^4$ coordinates, which in
the form language correspond to $dx^i$ and $*dx^i*$ (in fact, this is
true for all values of $N$ and $k$). Acting with these operators on
the state we found, we find 16 states, of the form $\prod_i
\delta(x^i)$ times any number of $dx^i$'s.

The $Spin(5)$ currents act only on the fermions, and not on the
coordinates of the moduli space. As described above, we have already
chosen a $U(1)_R \subset Spin(5)$ such that its charge corresponds (up
to a shift by $2Nk$) to the degree of a form. There is an
$SU(2)\subset Spin(5)$ which commutes with this $U(1)_R$, and the
decomposition of the adjoint $\bf{10}$ representation of $Spin(5)$
into $SU(2)\times U(1)_R$ representations is $\bf{3}_2 + \bf{1}_0 +
\bf{3}_0 + \bf{3}_{-2}$. The $\bf{1}_0$ component of the $Spin(5)$
charge is just the $U(1)_R$ generator, which may be written as $g_{ij}
dx^i * dx^j *$. The $\bf{3}_0$ components generate the $SU(2) \subset
Spin(5)$, and may be written as $g_{ij} (J^a)^j_k dx^i * dx^k *$,
where $J^a$ are the three covariantly constant complex structures of
the manifold. The $\bf{3}_2$ components are represented as a
multiplication by the three K\"ahler forms, each of which is a $(1,1)$
form with respect to one of the 3 covariantly constant complex
structures of the manifold. In our example of $N=1$ $k=1$, they may be
chosen to be $dx^1 \wedge dx^2 + dx^3 \wedge dx^4$, $dx^1 \wedge dx^3
- dx^2 \wedge dx^4$ and $dx^1\wedge dx^4 + dx^2 \wedge dx^3$. The
$\bf{3}_{-2}$ generators are the conjugates of these (by the star
operation).

The $SU(2)_R$ and $SU(2)_L$ currents are more complicated, since they
act also on the bosonic coordinates, and they correspond to
geometrical symmetries of the manifold. Obviously they commute with
$U(1)_R$, so they do not change the degree of a form. $SU(2)_R$ acts
only on the bosonic coordinates and not on the fermionic coordinates,
and in the simple case of $N=1$ $k=1$, it is a subgroup of the $SO(4)$
rotation symmetry of $\IR^4$, generated by $(x^i {\del \over \del x^j}
- x^j {\del \over \del x^i})$. $SU(2)_L$ is slightly more complicated,
since it acts also on the fermions -- its generators are similarly a
subgroup of $SO(4)$ but with an action also on the forms and not only
on their coefficient function.

Using these symmetry generators, it is easy to check that the 16
states we found above are in the $\bf(5,1)+\bf(4,2)+\bf(1,3)$
representation of $Spin(5)\times SU(2)_L$, and singlets of
$SU(2)_R$. This is also clear since the $\tilde{Q}$ generators are in
the $\bf(1,2,4)$ of $SU(2)_R\times SU(2)_L\times Spin(5)$, and the 16
states are generated by zero modes in this representation.

Other states in the superconformal representation of the free tensor
multiplet may be generated by acting with the other space-time
superconformal generators on the state we found. In this case the only
operators which do not vanish when acting on $F$ of \fgadol\ are
$P_i$, $H$ and $Q$, leading to various derivatives of the delta
function (all the states we describe here are eigenstates of $T$). The
operator $P_i$ acts by shifting $x^i$. Thus, it is clear that an
insertion of an operator at a position $x_0^i$ is described by
exchanging $\delta(x)$ above with $\delta(x-x_0)$.

Next, let us compute the 2-point functions of the chiral primary
states that we found. For instance, for the state $F$ described above,
the 2-point function will be given by $\int_{\IR^4} (e^{iHt} F(x_1)) 
\wedge *F(x_2)$, or explicitly
\eqn\twopoint{\langle F^\dagger(x_1,t) F(x_2,0) \rangle =
\int d^4x [e^{iRt (\del/\del x^i)^2} \delta^4(x-x_1)] \delta^4(x-x_2),}
where we denote $\delta^4(x) = \prod_{i=1}^4 \delta(x^i)$. The time
evolution in this case is very simple, and we find
\eqn\twopointexp{\langle F^\dagger(x_1,t) F(x_2,0) \rangle = -{1\over 
16\pi^2
R^2 t^2} e^{-i(x_1-x_2)^2/4Rt}.}
Up to normalization constants, this is exactly the expected result for
a chiral primary operator of dimension 2, as described in section
5. The other 16 states described above all have the same 2-point
functions with their conjugate states, while all off-diagonal 2-point
functions vanish.

\subsec{States and Correlation Functions for $N=1,k=2$ and $N=2,k=1$}

The next simplest example is either $N=1,k=2$ or $N=2,k=1$. In both of
these cases the moduli space is of the form $\IR^4 \times
\IR^4/\IZ_2$, which we will parametrize by coordinates $\tilde{x}^i$
and $v^i$ respectively, though the interpretation of the states is
quite different in the two cases\foot{As described in \S3.3, the
normalization we use for the Hamiltonian is also different in the two
cases. In this section we will use the normalization of the $N=2,k=1$
case.}. In either case, the parameter $\zeta$ is the usual blowup
parameter for the $\IR^4/\IZ_2$ singularity.

The formula \formula\ predicts two cohomology elements, one which is an
8-form and one which is a 6-form. We will denote the corresponding
states by $F_8$ and $F_6$. The 8-form is similar to the 4-form
described in the previous subsection, and we may construct the
corresponding chiral primary state even in the limit $\zeta \to 0$.
The corresponding form is $\delta^4(\tx) \delta^4(v) \prod_i d\tx^i
\prod_j dv^j$.  Again, we can translate this state in space using the
$P_i$, giving $\delta^4(\tx-x_0) \delta^4(v) \prod_i d\tx^i \prod_j
dv^j$. We can also translate the state in time using the evolution
operator $H$ in the quantum mechanics, which is simply given by $H =
{R\over 2} (\del_{\tx^i}^2 + \del_{v^j}^2)$.
This turns the state into
\eqn\qmstate{F_8(x_0,t) \simeq {1\over t^4} e^{-((\tx^i-x_0^i)^2 
+ (v^j)^2) / 2iRt} \prod_i d\tx^i \prod_j dv^j,}
up to ($R$-dependent) constants which we will ignore throughout.

It is again easy to compute 2-point functions of this state, and
we find
\eqn\coorfunc{\eqalign{ \langle F_8^\dagger(x_1,t_1) F_8(x_2,t_2) \rangle
&\propto \int d^4\tx d^4v {1\over t_1^4} {1\over t_2^4}
e^{((\tx^i-x_1^i)^2+(v^i)^2) / 2iRt_1}
e^{-((\tx^i-x_2^i)^2+(v^i)^2) / 2iRt_2} \cr & \propto
{1 \over (t_1-t_2)^4} e^{-i(x_1^i-x_2^i)^2 / 2R(t_1-t_2)}. \cr}}
Up to constants, this is exactly what we expect to find for a primary
operator of dimension $d=4$. In the case $N=2,k=1$ we identify this
operator with $(u_1^2)_2$, or equivalently $[(u_1)_1]^2$, while in the 
case
$N=1,k=2$ we identify this operator with $(u_2)_1$.
As above, we can also describe descendants of this state, fill out the
$Spin(5)$ representations, etc. 

Next, let us describe the 6-form state $F_6$. After we blow up the
$\IR^4/\IZ_2$ singularity, there is a non-trivial homology cycle
corresponding to the blown up ${\bf CP}^1$, and a corresponding
cohomology 2-form. In this special case, we can choose a
representative of the cohomology which is annihilated not only by $d$
but also by $d^\dagger = *d*$, restricted to the $v$ coordinates of
the moduli space, and we will denote it by ${\cal F}_2 = 
[{\cal F}_2(v)]_{ij}
dv^i \wedge dv^j$. This form is concentrated near the blown-up
singularity, and in the $\zeta \to 0$ limit gives a state that will be
concentrated at the singularity. The 6-form state that we find
corresponds to $F_6 = \delta^4(\tx) \prod_i d\tx^i \wedge {\cal F}_2$.

It is easy to compute the time evolution of this state, since 
${\cal F}_2$ is
annihilated by the Hamiltonian of the $v$ coordinates\foot{In 
the form language, $H\sim d
d^\dagger + d^\dagger d$.}. Thus, the evolution will be the same
as for $N=1,k=1$, with a trivial evolution in the $v$ space. As
before, this is consistent with this state having dimension 2, and we
identify this state with $(u_1)_2$ for $N=2,k=1$ and with $(u_1)_1$
for $N=1,k=2$. The analysis of the descendants of this state is
exactly as in the case of $N=1,k=1$ described above.

Note that
in the computations described in this section, we found
2-point functions that did not depend on $\zeta$, and agreed exactly
with our general predictions, described in \S5.2. This depended on
choosing particular representatives from each cohomology
class. Generally, as described above, there will be $\zeta$-dependence
and also different cohomology representatives will give different
2-point functions. But, the long time behavior will not depend on this.

\subsec{Comments on General $n$-point Functions}

In the previous sections we described the identification of some of
the states in the quantum mechanics corresponding to the space-time
theory for $P_-=N/R$. These states are the momentum $N$ modes of some
of the primary fields, $\phi(x^+,x^i,P_-=N/R)\vert 0\rangle$. As
described above, it is simple to compute 2-point functions of fields
using these states\foot{In some cases the primary field we find is the
non-singular part of a product of $l$ other primary fields, as
detailed in \S6.2, and then this 2-point function is also a special
case of a $2l$-point function, in which $l$ operators are at one point
and $l$ at another.}.

Obviously, we are also interested in computing higher $n$-point
functions. We will briefly outline here a procedure for such a
computation, although the details remain to be worked out \future. In
the space-time theory, we would like to compute an $n$-point function
of the form
\eqn\npt{\langle
\phi_1(x_1,t_1)\phi_2(x_2,t_2) \cdots \phi_n(x_n,t_n)\rangle.} 
In DLCQ, the
object that we may try and calculate is of the form
\eqn\npdlcq{\langle \phi_1\bigl(x^i_1,x^+_1,P_{-,1}={k_1\over
R}\bigr) \cdots \phi_n\bigl(x^i_n,x^+_n,P_{-,n}={k_n\over
R}\bigr)\rangle.}  To simplify the discussion we will limit ourselves
to correlation functions in which the operators with positive $P_-$
appear at smaller values of $x^+$ than the operators with negative
$P_-$. The prototypical problem is then to construct a state of
the form
\eqn\protstat{\phi_1\bigl(x^i_1,x^+_1,P_{-,1}={k_1\over R}\bigr)
\phi_2\bigl(x^i_2,x^+_2,P_{-,2}={k_2\over R}\bigr)\vert
0\rangle,} 
with $k_1,k_2 > 0$, 
which can then be used to compute 3-point or 4-point
functions\foot{Note that this state generally includes operators at
different ``DLCQ time'' $x^+$. This should be understood as involving
a time evolution of these states (using the Hamiltonian) to the same
$x^+$.}.

The state \protstat\ has longitudinal momentum $P_-=(k_1+k_2)/R$, so
it should appear in the quantum mechanics in the $N$-instanton moduli
space (where $N=k_1+k_2$). 
Without loss of generality we can assume that $x^+_2 < x^+_1$. We
start by constructing the state
$\phi_2\bigl(x^i_2,x^+_2,P_{-,2}={k_2/R}\bigr)\vert 0\rangle$. This is
a state in the quantum mechanics on the moduli space of $k_2$
instantons. At DLCQ time $x^+_2$ it is localized (as described above)
at the origin of the Higgs branch, and at some point $x^i_2$ in the
center of mass coordinates. We can propagate this state in the quantum
mechanics (using the Hamiltonian) up to the time $x^+_1$. Generally
the state will spread on the Higgs branch. Next, we would like to
transfer this state to the quantum mechanics on the $N$ instanton
moduli space, and to add the operator $\phi_1$.

This can be done in the following way. Within the $k_1+k_2$ instanton
moduli space we can focus on a submanifold $\cal W$ where $k_1$ of the
instantons have shrunk to zero size at some point $x^i_1$. The
submanifold $\cal W$ is roughly the moduli space of $k_2$ instantons.
These are the remaining instantons which can be in an arbitrary
instanton configuration. The space transverse to $\cal W$ is (locally)
roughly the moduli space of $k_1$ instantons, as going away from this
space corresponds to the $k_1$ instantons growing to a finite size
(or moving around). On $\cal W$ we can put the wave function that we
obtained above from the propagation of the $\phi_2$ state on the $k_2$
instanton moduli space. We can multiply this by a wave function on the
transverse space which is the wave function that corresponds to the
operator $\phi_1$. Since $\phi_1$ is a local operator, this wave
function is also localized on $\cal W$ (which is the origin of the
$k_1$ instanton moduli space). The resulting state corresponds to
\protstat\ at time $x^+_1$.

Once we have such a state we can propagate it in time, and then either
calculate its overlap with some state in the $N$ instanton moduli
space to obtain, say, a 3-point function, or we can again embed it
into an instanton moduli space of higher instanton number and
calculate a higher $n$-point function.

\vskip 1cm

\centerline{\bf Acknowledgments}\nobreak

We would like to thank J. Distler, R. Entin, N. Hitchin, S. Kachru,
S. Minwalla, G. Moore, J. Morgan, H. Nakajima, Y. Oz, E. Silverstein
and E. Witten for useful discussions. We are particularly indebted to
T. Banks and L. Susskind for collaboration in the early stages of this
work. OA was supported in part by DOE grant \#DE-FG02-96ER40559. The
work of MB was supported by NSF grant NSF PHY-9512835. The work of NS
was supported by DOE grant \#DE-FG02-90ER40542.

\listrefs

\end